\documentclass[10pt,twocolumn,letterpaper]{article}
\usepackage{latex8}
\usepackage{subfig}
\usepackage{graphicx}
\usepackage{array}
\usepackage{url}
\usepackage{eso-pic}

\AddToShipoutPicture*{\small \sffamily\raisebox{1.8cm}{\hspace{1.8cm}978-1-4244-4439-7/09/\$25.00 \copyright2009 IEEE}}

\begin{document}

\title{HYMAD: Hybrid DTN-MANET Routing for Dense and Highly Dynamic Wireless Networks}
\author{John Whitbeck \\
  Thal\`es Communications and \\
  UMPC Paris Universitas - LIP6
\and
Vania Conan \\
Thal\`es Communications
}

\maketitle
\thispagestyle{empty}

\begin{abstract}
  In this paper we propose HYMAD, a Hybrid DTN-MANET routing protocol
  which uses DTN between disjoint groups of nodes while using MANET
  routing within these groups. HYMAD is fully decentralized and only
  makes use of topological information exchanges between the nodes.
  We evaluate the scheme in simulation by replaying real life traces
  which exhibit this highly dynamic connectivity. The results show
  that HYMAD outperforms the multi-copy Spray-and-Wait DTN routing
  protocol it extends, both in terms of delivery ratio and delay, for
  any number of message copies.  Our conclusion is that such a Hybrid
  DTN-MANET approach offers a promising venue for the delivery of
  elastic data in mobile ad-hoc networks as it retains the resilience
  of a \textit{pure} DTN protocol while significantly improving
  performance.
\end{abstract}

\section{Introduction}
When transporting data through a wireless mobile ad-hoc network, the
Delay/Disruption-Tolerant Network (DTN)~\cite{dtn_fall_sigcomm}
paradigm uses node mobility as an advantage while compromising on
message delivery delays~\cite{GrossglauserTse2002}. Message forwarding
decisions are made on a \textit{per-encounter} basis, for example by
using utility functions based on aggregating statistics on node
meeting probabilities~\cite{lindgren03,daly07,LER}.  At any given
time, a node's vision of the network topology is limited to its
current neighbor. It does not have complete or even local knowledge of
the actual network topology as in the conventional Mobile Ad-hoc
Network (MANET) routing schemes. While this makes perfect sense in
extremely sparse networks~\cite{Burgess:2006,crawdad}, there are
situations where a highly mobile network is dense and sufficiently
well connected to provide end-to-end connectivity between a
significant subset of its nodes.

These nodes may even form small islands of stability. Using MANET
principles within such islands can bring great improvements. Indeed,
it considerably increases each node's information of its local
topology, thus leading to better forwarding decisions.  When high
mobility rates and more generally high link instabilities reduce route
life-times and threaten network-wide end-to-end connectivity, a MANET
routing protocol can still succeed locally even if it fails globally.

In this paper we propose HYMAD, a Hybrid DTN-MANET routing protocol.
HYMAD combines techniques from both traditional ad-hoc routing and DTN
approaches. HYMAD periodically scans for network topology changes and
builds temporary disjoint groups of connected nodes. Intra-group
delivery is performed by a conventional ad-hoc routing protocol and
inter-group delivery by a DTN protocol.

HYMAD constantly adapts to the dynamics of the wireless ad-hoc network
using only topological information. As in traditional ad-hoc routing,
no extra information on geographical location or social community
membership is required. It does not rely on a priori knowledge of
connectivity patterns or inter-meeting times. This makes HYMAD
amenable to implementation in a DTN stack or ad-hoc routing
protocol~\cite{JOTT06}. In a dense network, HYMAD can function
similarly to a traditional MANET protocol. In the other extreme case
of very sparse connectivity, each node is a group on its own and HYMAD behaves like
a classical DTN routing protocol. In any other intermediate case its
hybrid nature takes over.

We implemented the HYMAD hybrid approach with a self-stabilizing group
service~\cite{r_operators,DKP08} and the multi-copy Spray-and-Wait
protocol as the DTN routing scheme~\cite{spyro_sw}. We evaluated the
scheme by performing simulation runs on the Rollernet data
set~\cite{tournoux08_rollernet}, an example of a highly dynamic ad-hoc
network, and show that it brings substantial performance improvements
over \textit{pure} Spray-and-Wait.

In the next section, we further describe how our hybrid approach
positions itself compared to existing DTN and MANET approaches.  In
section~\ref{hymad}, we describe the HYMAD routing protocol
principles. We explain how nodes can agree on forming disjoint groups
and how such groups rather than individual nodes can be used as the
basis for DTN routing. We then evaluate the scheme on a real data set,
the Rollernet experiment, in section~\ref{results}. 
Finally we conclude our work in
section~\ref{conclusion}.

\section{Routing in a mobile wireless network}
\label{comparison}
\begin{figure}
  \centering
  \scalebox{0.8}{\includegraphics{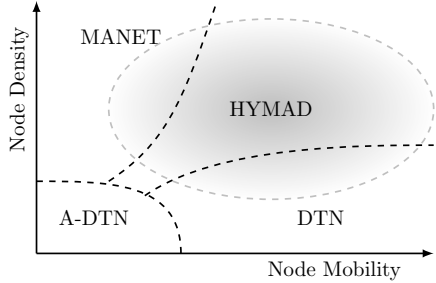}}
  \caption{Mobility vs Density: when different paradigms apply}
  \label{mod_density}
\end{figure}

Mobile wireless ad-hoc networks were first studied under the
assumptions of moderate node mobility and sufficient density to ensure
end-to-end connectivity. Both conditions are necessary for traditional
MANET approaches, be they proactive 
or reactive.

Let us characterize the various occurrences of mobile wireless networks
along the two main parameters of node density and node mobility. In
Fig.~\ref{mod_density}, which maps the different routing approaches on
the bi-dimensional mobile wireless network space, traditional MANET
routing appears in the top left corner.

When the density of nodes diminishes end-to-end connectivity can
disappear.  In such sparse networks nodes have very few, if any,
neighbors within their transmission ranges. The topology eventually
splits into several non-communicating connected components.  This is
typically the realm of Delay Tolerant Networking which one can further
subdivide in two~\cite{BorrelAmmar07}: the Assisted DTNs (A-DTN), in
case of low mobility of nodes, or Unassisted DTNs (U-DTN) where
mobility is high. The latter corresponds to traditional DTN scenarios.

Routing in A-DTNs typically involves special mobile
nodes, known as message ferries or data mules, which relay the messages
between the separate connected components
~\cite{ZA-ACMMOBIHOC2004,Shah2003}. The packet-switching method of
MANETs is replaced with a store-and-forward approach.

When the mobility in sparse networks increases, mobile nodes begin to
meet others. This is the traditional DTN scenario, where nodes forward
one or more copies of a given message until it reaches its
destination. There are many strategies for optimizing the forwarding
decision. The most straightforward approaches, such as Epidemic or
Spray-and-Wait~\cite{spyro_sw} do not require nodes to acquire
information on the others' positions, movements or trajectories.  More
elaborate schemes involve a utility function where each node collects
direct and indirect knowledge of other nodes' meeting
probabilities. They require a certain learning period to aggregate
statistics before making good forwarding decisions. For example,
Lindgren et al.~\cite{lindgren03} use past encounters to predict the
probability of meeting a node again while Daly et al.~\cite{daly07}
use local estimates of betweeness and similarity.

In dense networks, conventional MANET protocols start to break down
under high mobility down even if the network is almost always fully
connected. Indeed the sheer instability of the links would result in a
deluge of topology updates in the proactive case and \textit{route
  error} and new \textit{route requests} messages in the reactive
case. DTNs protocols on the other hand can handle high mobility
regardless of the density of the network. However by narrowly focusing
on per-encounter events, they ignore a lot of available
information. For example, simply asking nodes to regularly broadcast
a list of their neighbors would give all nodes a picture of its
two-hop neighborhood even under high mobility. Repeat this once and
everyone knows their three hop neighborhood. A node may therefore
have a topology ``knowledge horizon'' which determines how far into
the real topology a node can ``see''. The more extreme the mobility,
the shorter the ``horizon''.

The Hybrid DTN-MANET approach that we advocate in this paper aims at
filling the gap for efficient routing in highly connected and highly
mobile networks, which have so far, to the best of our knowledge,
received little attention.  Hybrid DTN-MANET routing, like the HYMAD
protocol that we describe below, combines the resilience of DTNs with
the greater knowledge of local network topology provided by a MANET
protocol. It adapts naturally to the dynamics of the network and its
applicability spans a large spectrum of the mobile wireless network
space.

\section{The HYMAD protocol}
\label{hymad}

\subsection{Overview}

The core idea in HYMAD is to use whole groups of nodes instead of
individual nodes as the focus of a DTN protocol. The analogy is
detailed as follows:

\renewcommand{\arraystretch}{1.2}
\begin{center}
\begin{tabular}{m{2.5cm}m{5.5cm}}
DTN & HYMAD \\
\hline
Node & Group of nodes \\
A node has message $m$ & One node in the group has message $m$ and all other nodes in the group know that. \\
Two nodes meet & Two disjoint groups become connected. \\
\hline
\end{tabular}
\end{center}

Each node $u$ regularly broadcasts a list detailing for each group
member $v$ including itself the following elements:

\begin{enumerate}
\item The minimal number of hops from $u$ to $v$.
\item A list of the messages held by $v$.
\item A bit indicating if $v$ is a \textit{border node}
  (i.e. in contact with other groups).
\end{enumerate}

The first two elements are necessary for the inter-group routing
protocol. The second one in particular allows a group to agree on what
messages it carries and which node (hereafter call the
message's custodian) specifically holds it. The last one enables use of an
intra-group distance vector routing. As in traditional distance vector
algorithms, the number of iterative broadcasts necessary for all
members of a group to agree on this information is equal to the
diameter of the group. 

HYMAD then uses a DTN protocol to transfer messages between
groups. The approach is generic and many existing DTN protocols could be employed.
In this paper, we use Spray-and-wait~\cite{spyro_sw} to forward messages
between disjoint groups. As in Spray-and-Wait, the source of a message
will create a certain number of copies of it. In HYMAD however, this
source node is part of a group and copies of the message will be
distributed among the adjacent groups instead of simply the nodes that
the source encounters. If a group has more than one copy, it will, in
turn, distribute extra copies to its other adjacent groups. If a group
has just one copy it will wait until encountering the destination's
group to transfer it. Once inside the destination's group, the
intra-group routing protocol delivers the message to the destination.

\subsection{Intra-group routing}
\label{intra_group}

\begin{figure}
  \centering
  \scalebox{0.8}{\includegraphics{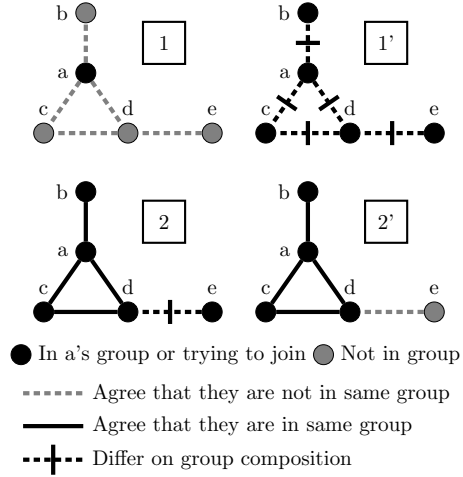}}
  \caption{Self-stabilizing groups: convergence in two iterations}
  \label{group_spread}
\end{figure}

In HYMAD, the intra-group routing is handled by a simple distance
vector algorithm. 

The nodes are dynamically grouped with a distributed network partitioning algorithm.
In our implementation, we chose to consider
\emph{diameter-constrained} groups. Groups will accept new members as
long as its diameter is less than a maximum diameter parameter
($D_{max}$). If a group's diameter expands due to internal link
failure, then some members are excluded to satisfy the diameter
constraint. Ducourthial et al.~\cite{DKP08} propose a
self-stabilizing, asynchronous distributed algorithm that achieves
this using an \mbox{$r-operator$} on a slightly modified distance
vector. This algorithm converges in $O(D_{max})$ iterations. The proof
of self-stabilization using asynchronous message passing can be found
in~\cite{r_operators}.

The main ideas behind group creation and modification are illustrated
in Fig.~\ref{group_spread} for a maximum diameter $D_{max}=2$. In the
first iteration, node $a$ begins by broadcasting the distance vector
$(a:0)$. Nodes $b$,$c$ and $d$ decide they want to join the group and
broadcast $(b:0,a:1)$, $(c:0,a:1)$ and $(d:0,a:1)$ respectively. After
receiving the broadcast from $d$, node $e$ also decides that it wants
to join the group and broadcasts $(e:0,d:1,a:2)$ (or
$(e:0,d:1,c:1,a:2)$ if $c$ spoke before $d$). In the second iteration,
$a$ now broadcasts $(a:0,b:1,c:1,d:1)$, $d$ realizes that the distance
between $b$ and $e$ is greater than $D_{max}$ and therefore chooses to
exclude $e$ from the group and broadcasts $(d:0,a:1,c:1,b:2)$. Finally
$e$ understands that it is not part of the group. After two
iterations, the group has stabilized on $a,b,c,d$. Now lets suppose
that at a later date the link between $a$ and $c$ goes down. Node $c$
now only receives the broadcasted distance vector $(d:0,a:1,b:2)$ from
$d$. It then understands that it is no longer part of the group. As is
obvious from this example, a given topology can result in very
different groups depending on the order in which the nodes speak.

In this paper, we used this algorithm in a proactive fashion where
each node node periodically runs the algorithm and broadcasts its
distance vector. Group composition therefore changes in reaction to
topology changes rather than routing needs.

\subsection{Inter-group routing}
\label{inter_group}

\textit{Border nodes} take care of most of the inter-group DTN
routing. Indeed, the periodic broadcast protocol described in
\ref{intra_group} puts them in the unique position of knowing both the
composition of two adjacent groups as well as the messages they
hold. \textit{Border nodes} may request the custodian of a message to
transfer one of more copies to it.

When a \textit{border node} learns that its group has acquired copies
of a message that a neighboring group does not possess, it has the
following choices:
\begin{itemize}
\item If the message's destination is in the neighboring group,
  request the message from its custodian and pass it on.
\item If its group has more than one copy of the message, request
  $min\left(1,\left\lfloor \frac{n_c}{n_b} \right\rfloor \right)$ copies from its
  custodian and pass them on.  ($n_c$ is the number of copies and
  $n_b$ the current number of border nodes in the group). The idea is
  to fairly spread a group's copies among its adjacent groups.
\item Otherwise do nothing
\end{itemize}

Conversely, when a \textit{border node} receives copies of a new
message from an adjacent group it can either:
\begin{itemize}
\item If the destination is in its group, forward the message to
  it using the inter-group routing protocol.
\item Otherwise, randomly select a group member to be the custodian
  for the copies. This is done to spread the burden over members of a
  group.
\end{itemize}

With this in place, when a node wants to send a message, it simply
adds it to its own list of messages. Through the intra-routing protocol,
in $O(D_{max})$ time, the group's \textit{border nodes} will become
aware of the new message and request copies to forward it on to the
adjacent groups.

\subsection{Discussion}
An internal link failure may cause a group to split into several
separate sub-groups due to its diameter increasing. In such a
situation, each sub-group only has a fraction of the messages of the
original group. Fortunately this is not really a problem. Firstly, the
intra-group protocol detailed in~\ref{intra_group} ensures that nodes
will update their message lists accordingly when removing nodes from
their group member lists, thereby preventing a sub-group from advertising
messages it does not have or any other such incoherences. Secondly,
certain subgroups may still be connected to each other. If either
sub-group has more than one copy of some messages, these will be
copied over the other sub-group. In any case HYMAD recovers gracefully
from group splits.

Choosing a diameter parameter for the group self-stabilization
algorithm involves a trade-off. On the one hand, increasing it will
expand each node's individual ``knowledge horizon'' of the actual
network topology. Fewer copies will cover a larger portion of the
network, which will naturally lead to faster delivery. On the other
hand, this comes at the cost of increasing the convergence time and
overhead of the group service. Ideally, the convergence speed should
be considerably faster than the speed of topology changes. In a sense,
extreme mobility may fundamentally limit a node's possible knowledge
of the network's topology. The increased overhead results from each
node regularly broadcasting a list of all messages in its
group. Larger groups mechanically lead to longer control messages. If
one is willing to incur the extra cost, the diameter can be set to
encompass the entire network. In such a situation, HYMAD resembles a
resilient MANET routing protocol using store-and-forward for message
transfers. Furthermore, in many mobile wireless scenarios, there are
underlying social dynamics at work which can sometimes drive nodes to
gather into loose communities. $D_{max}$ should be chosen so as to
allow the expected number of members per social group to neatly fit
into one self-stabilizing group.

\section{Results on Rollernet data}
\label{results}

\subsection{Methodology}

\begin{figure}[t]
  \centering
  \subfloat[Average node degree \label{avg_node_degree}]{\scalebox{0.55}{\includegraphics{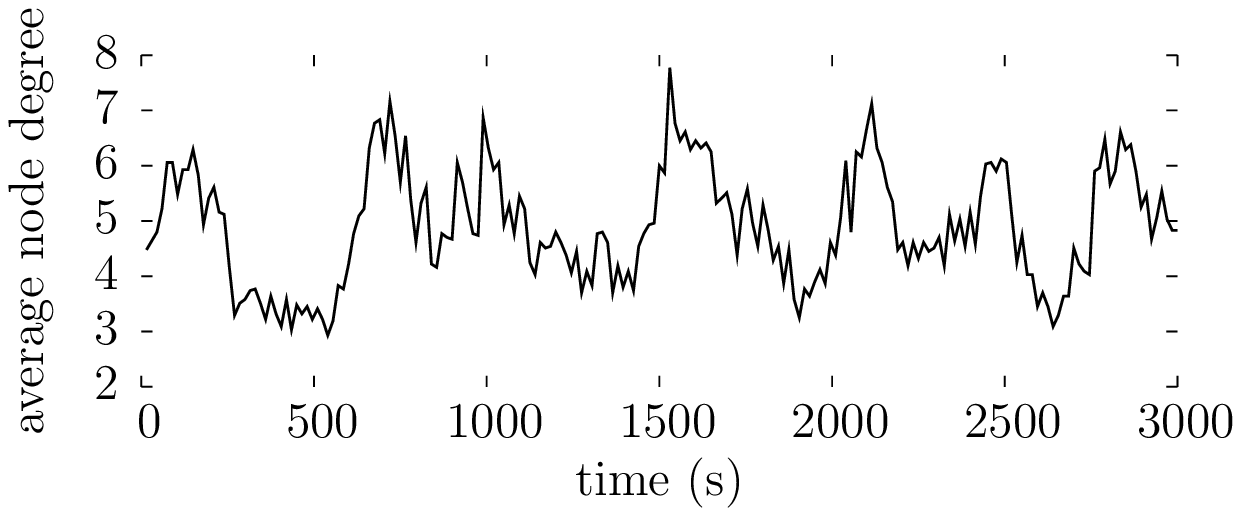}}} \\
  \subfloat[Number of connected components (ccs) \label{num_ccs}]{\scalebox{0.55}{\includegraphics{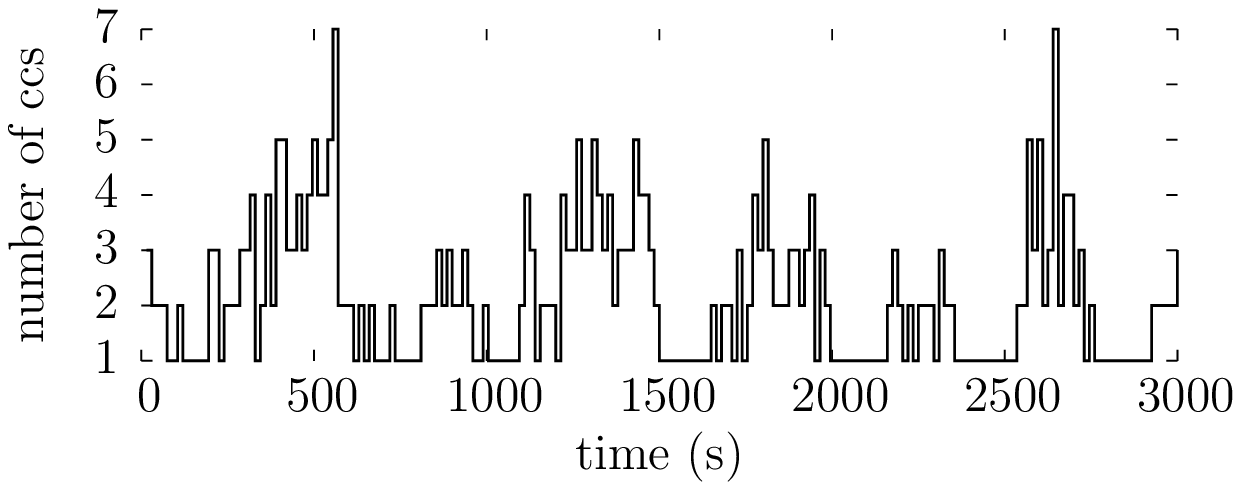}}} \\
  \caption{The accordion effect in Rollernet}
  \label{accordeon}
\end{figure}

\begin{figure}[t]
  \centering
  \scalebox{0.55}{\includegraphics{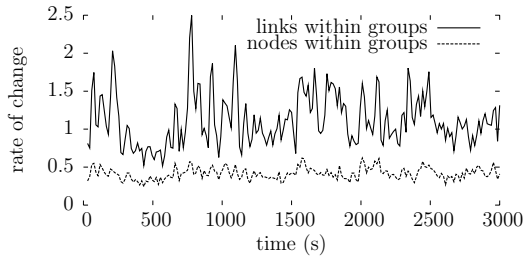}}
  \caption{Topology changes: average number of links within a given
    that appear or fail vs. number of nodes that leave or join that
    group}
  \label{group_stab}
\end{figure}

We evaluate HYMAD's performance on
Rollernet~\cite{tournoux08_rollernet}, a highly connected and
extremely mobile connectivity trace. The Rollernet experiment involved
equipping 62 participants of the regular Sunday afternoon rollerblading
tour through Paris with contact loggers (Intel iMotes). In order to
witness different behavior profiles the 62 bluetooth loggers were
distributed among groups of friends, members of rollerblading
associations and staff operators. In particular, one member of the
staff was instructed to remain behind the tour times while another
stayed in front for the entire duration of the experiment. This allows
us to get a rough sense of the relative geographic position of the
participants by looking at the connectivity graph. A snapshot of the
connectivity graph can be seen in Fig.~\ref{diffusion} and an
animation is available online~\cite{rollernet_youtube}.

The Rollernet trace is ideal for evaluating HYMAD. Indeed it exhibits
the following characteristics:
\begin{itemize}
\item \emph{High density:} Contrary to many DTN traces, Rollernet is
  \emph{not} sparse. A look at Fig.~\ref{avg_node_degree}, shows that
  the average node degree of the connectivity graph oscillates between
  2.9 and 7.8. The average for the whole tour is 4.8.
\item \emph{High mobility:} Everyone eventually meets everyone
  else. On average, each of the 62 nodes meets 56 others during the
  course of the tour. Additionally the topology evolves extremely
  quickly. The average lifetime of a given link is 26 seconds. The
  average lifetime of a shortest path between two nodes is 15.5
  seconds. Considering that the sampling period is 15 seconds, it
  follows that links are highly unstable and valid routes transient.
\item \emph{Accordion Effect:} This is an interesting consequence of
  the rollerblading context. The tour alternates between acceleration
  and deceleration phases in which the network topology respectively
  expands, leading to several separate connected component, and
  contracts, leading to a single connected
  component. Fig.~\ref{num_ccs} shows that the number of connected
  components varies between 1 and 7 (17 if counting isolated
  nodes). In fact, Figures \ref{avg_node_degree} and \ref{num_ccs} have
  roughly alternating phases.
\end{itemize}

We compare HYMAD to both Epidemic and regular Spray-and-Wait. Epidemic
provides an upper bound on achievable performance in terms of both
delay and delivery ratio while Spray-and-Wait provides a DTN
state-of-the-art comparison. We slightly adapted Spray-and-Wait to the
more connected context of Rollernet. A node no longer splits half of
its copies with the other nodes it meets, but instead splits its
copies equally among itself and its neighbors.

We chose to use $D_{max}=2$ for all our results because it ensures a
very fast convergence rate, keeps the overhead reasonable and seemed
to accurately reflect the size of separate connected components (small
groups of friends for example), particularly during the accelerating
phases. We also tested greater values of $D_{max}$, which yield , at
the cost of greater overhead, a small but noticeable improvement of
the delivery ratio.

The sampling period of the Rollernet traces is 15 seconds. We did not
try to extrapolate the events (link failures, new contact
opportunities, etc..) in the time between multiples of 15 seconds. We
also assume that 15 seconds is enough for a message to traverse any
connected component in Rollernet. Therefore, all our results on delays
when simulating protocols on top over the Rollernet traces will be in
multiples of 15 seconds.

\subsection{Performance}

\begin{figure}[t]
  \centering
  \subfloat[5 copies \label{cfs5}]{\scalebox{0.50}{\includegraphics{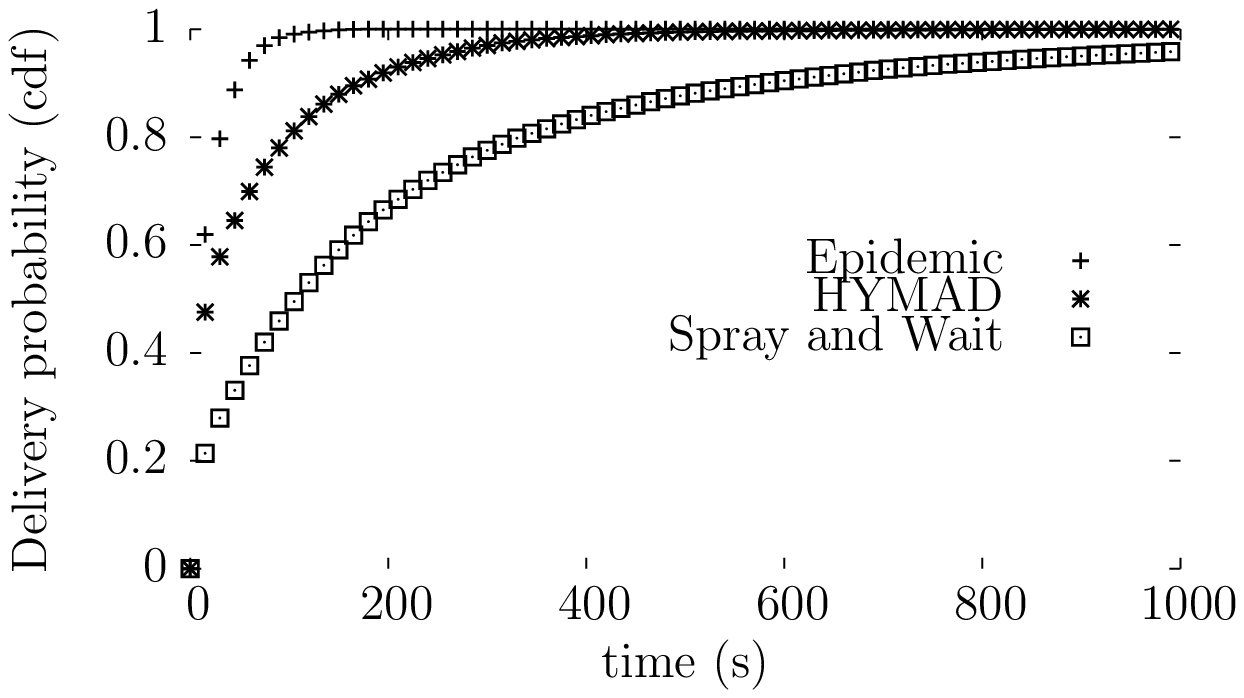}}} \\
  \subfloat[20 copies \label{cfs20}]{\scalebox{0.50}{\includegraphics{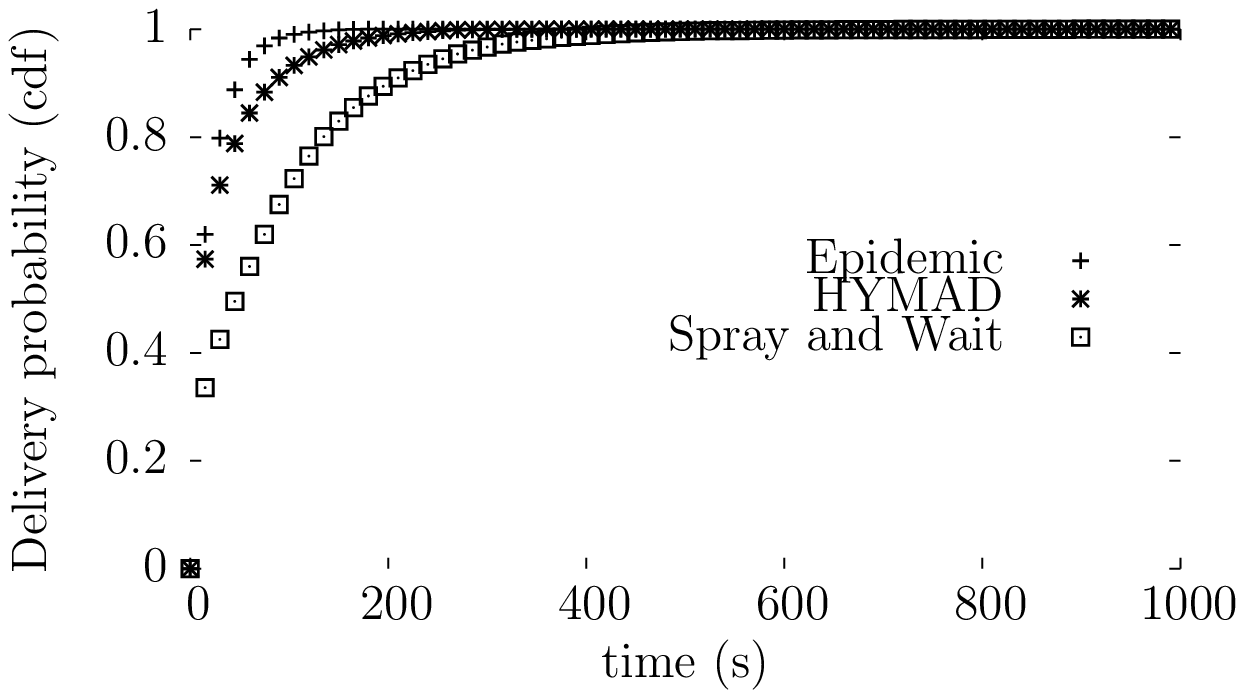}}}
  \caption{Comparison of delivery probabilities}
  \label{cfs}
\end{figure}

Extremely high link instability could mean one of two things. Either
nodes only briefly stay in the vicinity of one another or that nodes
may remain geographically close but that the link fails for other
reasons such as briefly moving out of transmission range or excessive
contention. We measured between each time step and from each node's
point of view, how many of its group members changed (number of new
members + number of excluded members) and how many links between
members of its group changed (either by appearing or
disappearing). The averages for all the nodes are shown in
Fig.~\ref{group_stab}. The composition of a given group appears much
more stable than the links among its members. This supports the idea
that small communities like groups of friends tend to stick together
during the tour and that link failures do not necessarily mean that
two nodes have clearly moved away from each other. Furthermore, the
rate of change of group composition, unlike most other metrics, seems
to smooth the accordion effect. This suggests that these groups are
indeed a good support for our hybrid approach.

To evaluate the performance of HYMAD we replayed the 3000 seconds of
the trace.  Every 15 seconds, during the first 2000 seconds, we
randomly selected 60 pairs of nodes which were instructed to send a
message to each other using Epidemic, HYMAD and Spray-and-Wait. This
averages results over both the connected and disconnected phases of
Rollernet. Figures \ref{cfs5} and \ref{cfs20} were obtained using the
aggregate data from 10 runs of this scenario with respectively 5 and
20 maximum number of copies for HYMAD and Spray-and-Wait. They compare
the cumulative distribution function of the delivery probability for
the three protocols. A few observations can be made:
\begin{itemize}
\item HYMAD clearly outperforms Spray-and-Wait in terms of delay and
  quickly achieves comparable performance with Epidemic.
\item With a low number of copies, HYMAD also outperforms
  Spray-and-Wait in terms of delivery ratio for reasons explained
  hereafter.
\item Predictably, performance increases with the number of
  copies. The maximum number of groups (including singletons) obtained
  at a given time is 29. Therefore, HYMAD with 20 copies will spray
  practically the entire network and therefore quickly and reliably
  reach the destination if in the same connected component as the
  source.
\end{itemize}

Spray-and-Wait's simple forwarding scheme performs very well
\emph{under the assumption of independent and identically distributed
  node mobility}~\cite{spyro_sw}. However this is absolutely
\emph{not} the case in Rollernet where groups of friends tend to stick
together. It is also usually \emph{not} the case in many real-world
situations where underlying social dynamics are often at work.

\begin{figure*}[t]
  \centering
  \includegraphics{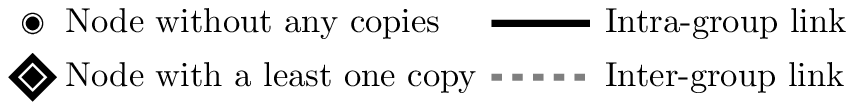} \\
  \subfloat[HYMAD: success within 15 seconds. \label{swg15}]{\includegraphics{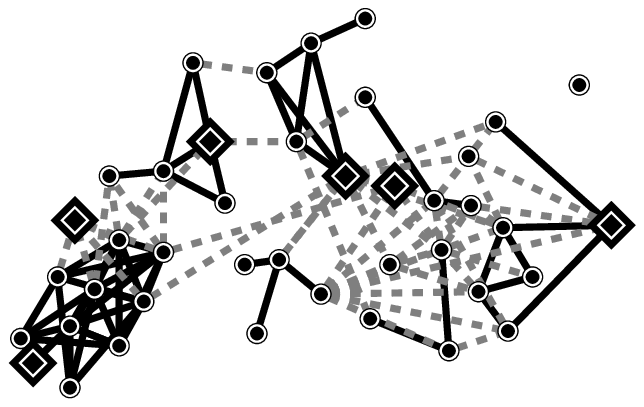}} \quad
  \subfloat[Spray-and-Wait: the copies stagnate around the source. It will take a total of 525 seconds to hit the destination. \label{sw15}]{\includegraphics{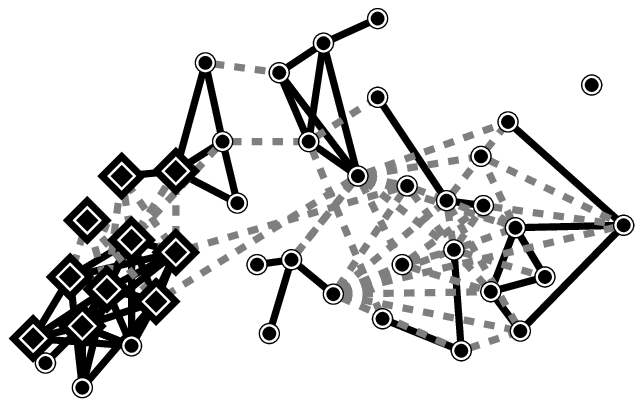}}
  \caption{Regular vs Hybrid Spray and Wait routing in Rollernet. (Partial view of the topology at t=15s)}
  \label{diffusion}
\end{figure*}

This can have a impact on performance. For example, when using just 5
copies, Spray-and-Wait simply fails to deliver about 5\% of messages
even after waiting for more than 15 minutes. Using 10 copies, the
average delay with Spray-and-Wait (133 seconds) is nearly three times
that of HYMAD (48 seconds). To further illustrate this point,
Fig~\ref{diffusion} compares the propagation of 10 copies after 15
seconds for HYMAD (Fig.~\ref{swg15}) and Spray-and-wait
(Fig.~\ref{sw15}). The rightmost node is the head of the rollerblading
tour. The bold lines represent intra-group links while the dashed gray
lines represent inter-group links. The nodes holding at least one copy
are represented by a diamond. In HYMAD's case, the destination is a
diamond meaning that our hybrid approach has delivered its message
within 15 seconds. On the other hand, the regular Spray-and-Wait
protocol distributed copies mainly within its own local group. These
nodes remain close to each other thus increase the delay. In this
particular case (Fig.~\ref{sw15}) it will take 525 seconds for a node
with a copy to meet the destination

\section{Conclusion and further work}
\label{conclusion}
In this paper we identified a new class of dense and highly mobile
networks not well addressed by conventional DTN or MANET approaches. We
proposed a new hybrid approach, HYMAD, that uses nodes' knowledge of
their local group topology to improve the performance of a simple DTN
protocol. In our case we used diameter-constrained groups along with
distance vector for intra-group routing and Spray-and-Wait for
inter-group routing. Simulations of our implementation in a dense and
highly mobile network show significant performance improvements over
regular Spray-and-Wait. Further work includes more comprehensive testing
against other real and synthetic mobility scenarios.

HYMAD is an example of a larger class of hybrid DTN-MANET routing
protocols which can handle a very wide spectrum of networks that
overlaps with those usually handled by either DTN or MANET.  We
believe that the first results that we obtained are encouraging for
further research in this direction. In particular, other more
elaborate DTN/MANET protocol pairs could conceivably be used for intra
and inter-group routing and would be worth exploring.

\section*{Acknowledgments}
This work has been partially supported by the RNRT project Airnet under contract
01205

\small
\bibliographystyle{latex8} 
\bibliography{adhoc_dtn}

\end{document}